\documentstyle[aps]{revtex}
\begin{document}
\newcommand{\beq}{\begin{equation}}
\newcommand{\eeq}{\end{equation}}
\newcommand{\beqn}{\begin{eqnarray}}
\newcommand{\eeqn}{\end{eqnarray}}
\newcommand{\bmath}{\begin{mathletters}}
\newcommand{\emath}{\end{mathletters}}
\title{Ferromagnetism from Undressing}
\author{J. E. Hirsch }
\address{Department of Physics, University of California, San Diego\\
La Jolla, CA 92093-0319}
 
\date{\today} 
\maketitle 
\begin{abstract}

We have recently  proposed that superconductivity may be understood as 
driven by the undressing of quasiparticles as the superconducting
state develops. Similarly we propose here that ferromagnetism in metals
may be understood as driven by the undressing of quasiparticles as the 
ferromagnetic state develops. In ferromagnets, the undressing is proposed
to occur due to the reduction in $bond$ $charge$ caused by spin polarization,
in contrast to superconductors where the undressing is proposed to occur
due to the reduction in $site$ $charge$ caused by (hole) pairing.
The undressing process manifests itself in the one and two-particle Green's 
functions as a transfer of spectral weight from high to low frequencies. 
Hence it should have universal observable consequences in one- and two-particle 
spectroscopies such as photoemission and optical absorption.

\end{abstract}

 \section{Introduction}

The concept of a quasiparticle is central to our understanding of
the physics of many-electron systems\cite{nozieres,schrieffer}. 
A quasiparticle is what remains
of a particle (electron) after taking into account its interaction with
surrounding particles. It may be understood as a particle carrying
with it a 'cloud' of other particles with which it interacts. 
This cloud can be visualized as 'clothing', or 'dressing', of the 
original particle, and naturally it will generally lead to an
increased effective mass of the quasiparticle.

In addition to the effective mass, another central element of the concept
of a quasiparticle is its 'weight'. Paradoxically, the 'weight' of a
quasiparticle is usually  the $inverse$ of its effective mass.
If a particle is heavily dressed by interactions, its effective mass
is large and its 'quasiparticle weight' is small. The 'quasiparticle
weight', $Z$, expresses how much of the particle still remains intact, 
with a well-defined energy versus momentum relation.
If $Z$ is small, the particle has lost most of its identity; by a
sum rule, this lost weight reappears elsewhere, in an incoherent
background. 

To recall the origin of the key connection between quasiparticle weight and its effective 
mass let us remember some basic concepts of many-body theory.
Consider the Green's function for a spin $\sigma$ electron in a 
many-body system
\beq
G(k,\tau)=<-T c_{k\sigma}(\tau)c^\dagger_{k\sigma}(0)> .
\eeq
G gives the probability amplitude to find an electron of momentum $k$
 at time
$\tau$ after it was created at  time $0$. Qualitatively, if the
electron does not interact with anything it will be found in its entirety
at time $\tau$ with the same momentum $k$ as when it was created
 at time 0. However if 
the electron interacts strongly with other degrees of freedom, we will find very 
little of it 
back at any later time. In that case the electron has lost its 'coherence', 
most or
all of it has been lost in an incoherent background. This is
why the Green's function in a many-body system is generally written as
\beq
G(k,\omega)=G_{coh}(k,\omega)+G'(k,\omega)
\eeq
where $G_{coh}$, the coherent part of the Green's function, represents the quasiparticle,
and $G'$ describes the incoherent background. This can be
understood as follows. The exact Green's function can be written as
\beq
G(k,\omega)=\frac{1}{\omega-\epsilon_k-\Sigma(k,\omega)}
\eeq
where $\Sigma$ is the self-energy, and $\epsilon_k$ is its kinetic enery 
measured from the chemical potential, inversely proportional to its mass. 
Assume for simplicity that the self-energy
has no $k$-dependence, and we have for its real part for small $\omega$
\beq
\Sigma_{re}(\omega)=\Sigma_{re}(0)+\omega\frac{\partial \Sigma_{re}}
{\partial \omega}
\eeq
$\Sigma_{re}(0)$ just renormalizes the chemical potential. From general phase
space arguments one knows that the imaginary part of $\Sigma$ goes to
zero as $\omega^2$ for small $\omega$. Hence we can separate the low frequency
real part of the Green's function and obtain
\beq
G(k,\omega)=\frac{1}{\omega(1-\frac{\partial \Sigma_{re}}{\partial \omega})
-\epsilon_k}+G'=\frac{Z}{\omega-Z\epsilon_k}+G'
\eeq
with
\beq
Z=\frac{1}{1-\frac{\partial \Sigma_{re}} {\partial \omega}}
\eeq
The term $G'$ contains the imaginary part of the self-energy and gives
rise to the incoherent contribution. Thus, Eq. (5) shows that the same
factor $Z$, the wave function renormalization factor, 
determines the quasiparticle spectral weight and the
effective mass renormalization.

This deep connection between quasiparticle weight and its effective mass is
well known. However it has not been stressed in the recent literature. For
example, there is a vast recent literature on the phenomenon of colossal
magnetoresistance in manganites, where the transition to the ferromagnetic
state is thought to be accompanied by a reduction in the carrier's effective mass.
Yet there has been no discussion to our knowledge of any corresponding 
change in the quasiparticle weight. Here we will make such a connection,
for the manganites and for ferromagnetic metals in general.

We have pointed out in the past that within a class of model Hamiltonians  
both superconductivity\cite{super}
and ferromagnetism\cite{ferro}
may be understood as driven by a lowering of the carrier's effective mass as
the ordered state develops, or equivalently a lowering of the carrier's
kinetic energy, and proposed that this common aspect of the physics
may be essential to both phenomena\cite{both}. Only recently however  
have we focused on the fundamental
connection between the lowering of effective mass and the corresponding
expected increase in the quasiparticle weight, or 'undressing', for the case of
superconductivity(\cite{undr}, hereafter referred to as I). 
This connection was brought to the limelight by the beautiful experimental
results and insightful analysis of Ding et al\cite{ding} on
photoemission in cuprates, as well as experimental work by
Feng et al\cite{shen} and Basov et al\cite{basov}. Here we make this connection
for the case of ferromagnetism. It leads to a remarkably simple picture
of metallic ferromagnetism, and to the understanding that  both ferromagnetism 
and superconductivity may be driven by the same
physical principle: undressing.

\section{The Physics}

Consider the process of creating an electron of spin $\uparrow$ at site
$i$. Imagine there is an electron of spin down at the bond connecting
site $i$ to a neighboring site $j$, as shown in Fig. 1. The $\downarrow$
electron is in state $|0>$, its ground state, in the absence of 
occupation of site $i$. The strong Coulomb repulsion between like charges
will affect the state of that bond charge.
When the $\uparrow$ electron is created
at site $i$, the $\downarrow$ electron will make a transition to one of the bond
states $|1^l>$, the eigenstates of the bond $\downarrow$ electron
in the presence of an $\uparrow$ electron at site $i$. Let the 
ground state of that manifold be $|1>\equiv |1^0>$, and we denote
by
\beq
S=<0|1>
\eeq
the overlap matrix element of the ground states of the $\downarrow$
bond electron in the absence and in the presence of an $\uparrow$ electron
at site $i$. We can express this mathematically as
\beq
c_{i\uparrow}^\dagger |0>|0>=|\uparrow>|0>=\sum_l|\uparrow>
|1^l><1^l|0>=|\uparrow>|1>S+\sum_{l\neq 0}|\uparrow>
|1^l><1^l|0>.
\eeq
Here, the first ket denotes the electronic state of the site $i$,
and the second ket denotes the state of the $\downarrow$ electron
at the bond. The first term in Eq. (8), where the $\downarrow$
electron ends up in its ground state $|1>$, represents
a coherent process that preserves the phase of the wave function. 
It is a 'diagonal transition' in the language of small polaron theory\cite{holstein}.
If instead the $\downarrow$ electron at the bond ends up in an excited
state, it represents an incoherent process (non-diagonal transition).
The overlap matrix element $S$ represents what fraction of the
$\uparrow$ electron created at site $i$ remains coherent,
and gives rise to its reduced quasiparticle  'weight'. If $S$ is very small it means most 
of the effect
of the $\uparrow$ electron creation has been dissipated in
incoherent processes that left the 'background' $\downarrow$ electron
in excited states.

If instead we  create the $\uparrow$ electron at site $i$ when there is
no  electron at bond $ij$ we have simply
\beq
c_{i\uparrow}^\dagger |0>=|\uparrow>
\eeq
and no incoherent piece is generated, because there was no
'background' degree of freedom to excite; hence, the weight of the
quasiparticle is $1$ in this case.

More generally, we could instead have an $\uparrow$ electron on the bond,
or both $\uparrow$ and $\downarrow$ electrons. We argue that Eq. (8) still applies,
with the second ket denoting the state of the total charge at the bond.
If there is no charge at the bond, Eq. (9) instead applies. We define
then a quasiparticle operator $\tilde{c}_{i\uparrow}$ through the
relation
\beq
c_{i\uparrow}^\dagger=[1-
(1-S)\frac{(\tilde{n}_{ij\uparrow}+\tilde{n}_{ij\downarrow})}{2}
]\tilde{c}_{i\uparrow}^\dagger
\eeq
where $\tilde{n}_{ij\sigma}$ is the bond occupation number ($0$ or $1$) of the
spin $\sigma$ electron. Eq. (10) is only the coherent part of the electron
operator, as it does not generate the second part of Eq. (8)
(we use the same operator notation on the left for simplicity only). Eq. (10)
 restates Eqs. (8) and (9), that the weight of $c_{i\uparrow}^\dagger$ is
$1$ if the neighboring bond is unoccupied, and it is maximally reduced to
$S<1$ if the
neighboring bond is occupied by  spin $\uparrow$ and $\downarrow$ electrons.

The physics resulting from these  equations is shown schematically
in Fig. 2. As in I, we use an 'independent boson model' with an Einstein 
oscillator\cite{mahan} to 
describe the coupling of the electron  at site $i$  to the 
excited states of the charge (in this case the bond charge) shown in Fig. 1.
The coherent part of the
spectral function (quasiparticle peak, labeled q.p.) arises from
the ground-state to ground-state transition of the oscillator when the
electron is created at the site, and its height is the quasiparticle
weight $Z$. $Z$ increases as the bond charge decreases, and
correspondingly weight in the spectral function shifts from the 
incoherent part to the quasiparticle peak. As we will show in the
next section, the bond charge decreases when spin polarization develops.

If instead of focusing on the bond charge we were to focus on the site charge
the equation analogous to (10) is
\beq
c_{i\uparrow}^\dagger=[1-(1-S)\tilde{n}_{i\downarrow}]\tilde{c}_{i\uparrow}^\dagger
\eeq
where $\tilde{n}_{i\downarrow}$ is the site charge occupation.
It was shown in I that Eq. (11) leads to superconductivity through undressing.

Next we wish to express the bond charge in terms of electron
operators. We use the operator representation
\beq
\tilde{n}_{ij\sigma}=\tilde{c}_{i\sigma}^\dagger \tilde{c}_{j\sigma}
+\tilde{c}_{j\sigma}^\dagger \tilde{c}_{i\sigma}
\eeq
which has eigenvalue $1$ operating on the low energy bonding state
\beq
\frac{|\sigma>_i|0>_j+|0>_i|\sigma>_j}{\sqrt{2}}
\eeq
and zero if the bonding state is empty. Eq. (10) is then
\beq
c_{i\uparrow}^\dagger=[1-(1-S)\frac{1}{2}\sum_\sigma
(\tilde{c}_{i\sigma}^\dagger \tilde{c}_{j\sigma}
+\tilde{c}_{j\sigma}^\dagger \tilde{c}_{i\sigma})
]\tilde{c}_{i\uparrow}^\dagger
\eeq
Eq. (14) is the analog, for the effect of creating the $\uparrow$ electron
on site $i$ on the neighboring bond charge, to Eq. (11) for the effect
of creating the $\uparrow$ electron on site $i$  
on the site charge. Note an important difference: under a particle-hole
transformation
\beq
c_{i\sigma}^\dagger\rightarrow (-1)^ic_{i\sigma}
\eeq
(on a bipartite lattice) Eq. (14) is invariant, while Eq. (11) changes to
\beq
c_{i\uparrow}^\dagger=[S+(1-S)\tilde{n}_{i\downarrow}]\tilde{c}_{i\uparrow}^\dagger
\eeq
Eq. (11) implies that increasing $electron$ site concentration leads to 
increased $dressing$ of electrons, and conversely Eq. (16) implies that
increasing $hole$ site concentration leads to $undressing$ of holes.
Instead, Eq. (14) and its identical form in hole representation imply
that increasing bond occupation leads to increased dressing, both for
electrons and holes. This difference between the dressing effects of site
and bond charges lies at the root of the difference between
superconductivity and ferromagnetism.

Finally, we consider a d-dimensional hypercubic lattice and add the
contributions from all the bonds connecting to a given site,
and Eq. (14) becomes
\beq
c_{i\sigma}^\dagger=
[1-(1-S)\frac{1}{2}\sum_{\delta,\sigma'}
(\tilde{c}_{i\sigma '}^\dagger \tilde{c}_{i+\delta \sigma '}
+\tilde{c}_{j\sigma '}^\dagger \tilde{c}_{i\sigma '})
]\tilde{c}_{i\sigma }^\dagger   .
\eeq
We explore its consequences
 in the next sections.

\section{Quasiparticle Hamiltonian}
Consider the kinetic energy operator on a lattice
\beq
H_{kin}=-\sum_{i,j,\sigma} t_{ij}c_{i\sigma}^\dagger c_{j\sigma}
\eeq
Replacing the bare electron operators in Eq. (18) by the quasiparticle operators
Eq. (17) yields the low energy effective Hamiltonian for quasiparticles
\beqn
H_{kin}&=&-\sum_{i,j,\sigma} t_{ij}\times \nonumber \\
&[&1-\frac{(1-S)}{2}\sum_{\delta,\sigma'}
(\tilde{c}_{i\sigma '}^\dagger \tilde{c}_{i+\delta \sigma '}
+\tilde{c}_{i+\delta \sigma '}^\dagger \tilde{c}_{i\sigma '})]
\tilde{c}_{i\sigma}^\dagger \tilde{c}_{j\sigma}
[1-\frac{(1-S)}{2}\sum_{\delta,\sigma'}
(\tilde{c}_{j\sigma '}^\dagger \tilde{c}_{j+\delta \sigma '}
+\tilde{c}_{j+\delta \sigma '}^\dagger \tilde{c}_{j\sigma '})]
\eeqn
In expanding this expression we will ignore terms involving more than 
two centers for simplicity, as well as terms with more than four fermion
operators. The latter can certainly be rigurously justified if the
electron (or hole) density is low.  Eq. (19) then becomes
\bmath
\beq
H_{kin}=-\sum_{i,j,\sigma} t_{ij}\tilde{c}_{i\sigma}^\dagger \tilde{c}_{j\sigma}
+(1-S)\sum_{<i,j>}t_{ij}(\sum_\sigma
\tilde{c}_{i\sigma}^\dagger \tilde{c}_{j\sigma} +h.c.)^2
\eeq
which can also be written as
\beq
H_{kin}=-\sum_{i,j,\sigma} t_{ij}\tilde{c}_{i\sigma}^\dagger \tilde{c}_{j\sigma}
+2(1-S)\sum_{i,j,\sigma\sigma'}t_{ij}
\tilde{c}_{i\sigma}^\dagger \tilde{c}_{j\sigma}
\tilde{c}_{j\sigma '}^\dagger \tilde{c}_{i\sigma '}
+2(1-S)\sum_{i,j,\sigma\sigma'}t_{ij}
\tilde{c}_{i\sigma}^\dagger \tilde{c}_{j\sigma}
\tilde{c}_{i\sigma '}^\dagger \tilde{c}_{j\sigma '}
\eeq
\emath
In the form Eq. (20a), the interaction term generated can be simply understood
as bond-charge Coulomb repulsion\cite{schrieffer2,campbell,ferro4}. In the form
Eq. (20b) it can be seen that the two interaction terms are
 precisely of the same form as
the exchange and pair hopping terms that result from considering
off-diagonal matrix elements of the Coulomb interaction in a tight
binding representation\cite{hubbard}:
\bmath
\beq
J_{ij}=\int d^3r d^3r' \phi^*_i(r)\phi^*_j(r')\frac{e^2}{|r-r'|}\phi_i(r')\phi_j(r)
\eeq
\beq
J'_{ij}=\int d^3r d^3r' \phi^*_i(r)\phi^*_i(r')\frac{e^2}{|r-r'|}\phi_j(r')\phi_j(r)
\eeq
\emath
In general, we will have $J=J'$ from Eq. (21) if the wavefunctions can
be assumed to be real. Eq. (20) implies
\beq
J_{ij}=J'_{ij}=2t_{ij}(1-S)
\eeq
Supplementing the kinetic energy with an on-site Coulomb repulsion
leads to the low energy effective Hamiltonian
\beqn
H &=& -\sum_{<ij>,\sigma} t_{ij}(\tilde{c}_{i\sigma}^\dagger \tilde{c}_{j\sigma}+h.c.)
+U\sum_i\tilde{n}_{i\uparrow}\tilde{n}_{i\downarrow} \nonumber \\
&+&\sum_{i,j,\sigma\sigma'}J_{ij}
\tilde{c}_{i\sigma}^\dagger \tilde{c}_{j\sigma}
\tilde{c}_{j\sigma '}^\dagger \tilde{c}_{i\sigma '}
+\sum_{i,j,\sigma\sigma'}J'_{ij}
\tilde{c}_{i\sigma}^\dagger \tilde{c}_{j\sigma}
\tilde{c}_{i\sigma '}^\dagger \tilde{c}_{j\sigma '}
\eeqn
to describe the dynamics of the quasiparticles.

We have extensively studied the properties of this Hamiltonian for nearest
neighbor hopping $t_{ij}=t$ and interactions $J_{ij}=J$, $J'_{ij}=J'$, in
particular for the cases $J'=0$\cite{ferro} and $J'=J$\cite{ferro4}. 
The 'exchange term' involving $J$ can also be written as
\bmath
\beq
H_J=-2J\sum_{<ij>} (\vec{S}_i\cdot \vec{S}_j +\frac{1}{4} n_in_j)
\eeq
with
\beq
(\vec{S_i})_\alpha =\frac{1}{2}(c_{i\uparrow}^\dagger,c_{i\downarrow}^\dagger)
\sigma_\alpha \pmatrix{c_{i\uparrow} \cr c_{i\downarrow}}
\eeq
\emath
and $\sigma_\alpha$ a Pauli matrix ($\alpha=x,y,z$). In the form Eq. 
(24) it looks like a 'Heisenberg exchange' term. However, as
emphasized earlier\cite{ferro,ferro4}, the origin of ferromagnetism here is $not$ 
quantum-mechanical exchange of localized spins, as in Heisenberg's case. 
The combination of the $J$ and $J'$ terms in the form Eq. (20a) displays the
origin of these interactions as  bond-charge Coulomb repulsion. Ferromagnetism in this
model is driven by reduction of bond-charge Coulomb repulsion as spin polarization
 develops and accompanying
kinetic energy lowering, rather than quantum-mechanical exchange.

The properties of the model Eq. (23) for nearest neighbor hoppings and interactions
\beq
H=-t\sum_{<ij>,\sigma}(\tilde{c}_{i\sigma}^\dagger \tilde{c}_{j\sigma}+h.c.)
+U\sum_i\tilde{n}_{i\uparrow}\tilde{n}_{i\downarrow}
+J\sum_{i,j,\sigma\sigma'}
\tilde{c}_{i\sigma}^\dagger \tilde{c}_{j\sigma}
\tilde{c}_{j\sigma '}^\dagger \tilde{c}_{i\sigma '}
+J'\sum_{i,j,\sigma\sigma'}
\tilde{c}_{i\sigma}^\dagger \tilde{c}_{i,\sigma'}^\dagger
\tilde{c}_{j\sigma '} \tilde{c}_{j\sigma }
\eeq
are similar for different values of $J'/J$\cite{ferro,ferro4,ferro5}.
In particular, within mean field theory and for a model with constant
density of states, the conditions on the parameters for ferromagnetism to occur are
\bmath
\beq
j > \frac{1-u}{2-m^2-(1-n)^2}
\eeq
\beq
j > \frac{1-u}{\frac{5}{3}-\frac{3}{2}m^2-\frac{1}{2}(1-n)^2}
\eeq
\emath
for $J'=0$ and for $J'=J$ respectively, with
\bmath
\beq
u=U/D
\eeq
\beq
j=zJ/D
\eeq
\beq
j'=zJ'/D
\eeq
\emath
Here, $m$ is the magnetization per site, $n$ the total occupation per site,
$z$ the number of nearest neighbors to a site
and $D=2zt$  the bare bandwidth.
In particular, for the half-filled band ($n=1$) the condition for
full spin polarization ($m=n$) is
\beq
j=\frac{J}{2t}>1-u
\eeq
in both cases. For $U=0$, this condition is achieved in the limit
$S\rightarrow 0$ according to Eq. (22), while for increasingly
larger $U$ smaller values of $J$ are required, and hence larger
values of $S$ are sufficient. Exact diagonalization studies of
the Hamiltonian Eq. (25) show that the mean field conditions Eq. (26) are
qualitatively correct and reasonably accurate particularly
for the half-filled band and not too large values of $U$\cite{exact}.

In what follows we will for simplicity consider the model with $J$ only:
\beq
H=-t\sum_{<ij>,\sigma}(\tilde{c}_{i\sigma}^\dagger \tilde{c}_{j\sigma}+h.c.)
+U\sum_i\tilde{n}_{i\uparrow}\tilde{n}_{i\downarrow}
+J\sum_{i,j,\sigma\sigma'}
\tilde{c}_{i\sigma}^\dagger \tilde{c}_{j\sigma}
\tilde{c}_{j\sigma '}^\dagger \tilde{c}_{i\sigma '}
\eeq
One argument for dropping the $J'$ term is that its importance is
suppressed due to on-site Coulomb repulsion. However, in treating the
Hamiltonian in mean field theory this effect may not be properly
taken into account.
The effective kinetic energy that results from Eq. (29) within 
mean field theory is\cite{ferro}
\bmath
\beq
\epsilon_{k\sigma}=(1-2j (I_\uparrow+I_\downarrow))\epsilon_k
\eeq
with
\beq
I_\sigma = <\tilde{c}_{i\sigma}^\dagger \tilde{c}_{i+\delta,\sigma}>
\eeq
\emath
(one half of) the average bond charge for spin $\sigma$. 
$\epsilon_k$ is the Fourier transform of the bare hopping
amplitude $t_{ij}$. For a model with a constant density of states,
the average bond charge at zero temperature is given by\cite{ferro}
\beq
I_\uparrow +I_\downarrow\equiv I=\frac{1-m^2-(1-n)^2}{2}
\eeq
so that it decreases with increasing magnetization, as expected.

The properties of the model are simplest in
the half-filled band case, and we will restrict ourselves to that case
in what follows. For that case,
\beq
I_\uparrow=I_\downarrow=\frac{I}{2}
\eeq
even in the presence of spin polarization. As the temperature is
lowered below $T_c$ and ferromagnetism develops, the average bond 
charge
decreases for both majority and minority spins\cite{ferro}, and the effective
bandwidth 
\beq
D_{eff}=(1-2jI)D
\eeq
broadens. 
In the normal state, as the temperature decreases the
bond charge occupation Eq. (30b) increases and hence the bandwidth
narrows. The temperature and magnetization dependence of the bond charge
leads to a variety of interesting properties of the mean field
solution of this model that are not found in the Stoner model,
i.e. the mean field solution of the repulsive Hubbard model, and 
that describe experimental observations, as discussed in the 
references\cite{ferro}.

What is the significance of having derived the Hamiltonians Eq. (23)
or Eq. (25)
in this new way? Twofold. First, if one assumes that the interactions
$J$ and $J'$ in Eq. (23) arise from off-diagonal matrix elements of
the Coulomb interaction as given by Eq. (21), their value is expected
to be rather small. This is because one has to use properly orthogonalized
atomic orbitals in Eq. (21). When the Mulliken approximation\cite{mulliken} holds,
which is usually the case,
off-diagonal matrix elements such as $J$ and $J'$ are very small for
orthogonalized orbitals, and this
is the justification for the 'zero differential overlap' approximation
in quantum chemistry\cite{parr}. However, there is a  more fundamental reason
why the present derivation of the Hamiltonian Eq. (23) is more satisfactory
than the one using the Coulomb matrix elements argument. This is 
discussed in the next section.

\section{Ferromagnetism and spectral weight transfer}

The single particle Green's function (for spin $\uparrow$ electrons) is given by
\beq
G_{ij}(\tau)=<-Tc_{i\uparrow}(\tau)c_{j\uparrow}^\dagger(0)>=
G_{ij}^{coh}(\tau)+G_{ij}^{incoh}(\tau)
\eeq
with $T$ the time ordering operator.
The coherent and incoherent parts of the Green's function arise from the
first and other terms in Eq. (8) respectively. For the coherent part,
we replace the electron operators in terms of quasiparticle operators and obtain
\beq
G_{ij}^{coh}(\tau)=<-T[1-\frac{(1-S)}{2}\tilde{n}_{bond,i}(\tau)]
\tilde{c}_{i\uparrow}(\tau)\tilde{c}^\dagger_{i\downarrow}(0)
[1-\frac{(1-S)}{2}\tilde{n}_{bond,j}(0)]>
\eeq
where $n_{bond,i}$ represents the bond charge adjacent to site $i$. A mean field
decoupling leads to
\beq
G_{ij}^{coh}(\tau)=[1-(1-S)<\tilde{n}_{bond}>]
<-T\tilde{c}_{i\uparrow}(\tau)\tilde{c}^\dagger_{i\downarrow}(0)>
\equiv Z<-T\tilde{c}_{i\uparrow}(\tau)\tilde{c}^\dagger_{i\downarrow}(0)>
\eeq
Equation (36) defines the quasiparticle weight $Z$. We take for the average bond charge
\bmath
\beq
<\tilde{n}_{bond}>=\sum_\sigma <c^\dagger_{i\sigma}c_{j\sigma}+h.c.>
=2(I_\uparrow +I_\downarrow) =2I
\eeq
 which can be written as\cite{ferro}
\beq
I=\int_{-D/2}^{D/2} d\epsilon g(\epsilon)(-\frac{\epsilon}{D/2})
[f(\epsilon_\uparrow (\epsilon)+f(\epsilon_\downarrow (\epsilon)]\equiv I(T,m)
\eeq
\emath
and is a function of temperature and magnetization. Here,
 $\epsilon_\sigma(\epsilon)$ are the quasiparticle energies and $g(\epsilon)$ the 
density of states. Hence the quasiparticle weight is simply
\beq
Z=Z(T,m)=1-2(1-S)I(T,m)=1-2jI(T,m)
\eeq
and will depend on temperature and magnetization through the temperature and
magnetization dependence of the bond charge.
Note that for a more general case with a non-half-filled band, one would have different
quasiparticle weights $Z_\sigma$ for spin up and down electrons in the spin-polarized
state. Such a situation is also easily treated within this framework.

We consider the mean field solution of the model
Eq. (29). The quasiparticle energies are given by\cite{ferro}
\beq
\epsilon_{k\sigma}=
\epsilon_\sigma(\epsilon_k)=(1-2jI)\epsilon_k-\sigma\frac{U+Jz}{2}m-\mu
\eeq
where the magnetization $m$ and chemical potential $\mu$ are determined
by the conditions
\bmath
\beq
m=\int_{-D/2}^{D/2} d\epsilon g(\epsilon)
[f(\epsilon_\uparrow (\epsilon)-f(\epsilon_\downarrow (\epsilon)]
\eeq
\beq
n=\int_{-D/2}^{D/2} d\epsilon g(\epsilon)
[f(\epsilon_\uparrow (\epsilon)+f(\epsilon_\downarrow (\epsilon)]
\eeq
\emath
with $n$ the carrier concentration. From Eq. (39), the effective bandwidth is
given by
\beq
D_{eff}=(1-2jI)D=[1-2(1-S)I]D
\eeq
where for the last equality we have used Eqs. (22) and (27b).
The effective mass is given by
\beq
\frac{m^*}{m_0}=\frac{1}{1-2jI}
\eeq
where $m_0$ is the bare mass determined by the bare hopping amplitude in
Eq. (18).
From Eqs. (38) and (42) we have simply
\beq
\frac{m^*}{m}=\frac{1}{Z(T,m)} 
\eeq
as expected from Eq. (5).

The behavior of $I(T,m)$ is discussed in the references\cite{ferro}. In Fig. 3 we
reproduce a representative case. As the temperature is
lowered in the normal state $I(T,m)$ increases, and it decreases again as
spin polarization develops. Correspondingly, the bandwidth $D_{eff}$
decreases in the normal state upon cooling and expands again as the
ordered state develops; similarly the quasiparticle weight $Z$ decreases
above $T_c$ as $T$ decreases, and increases as spin polarization develops.
If no magnetization were to develop, for the parameters in Fig. 3
the effective bandwidth would shrink to zero as the temperature
goes to zero.

The coherent part of the spectral function is given by
\beq
A_{\sigma coh}(k,\omega)=
-\frac{1}{\pi}ImG_{coh}(k,\omega+i\delta)=Z\delta(\omega-\epsilon_{k\sigma})
\eeq
We can model the full spectral function by assuming a harmonic oscillator
spectrum of frequency $\omega_0$ associated with the bond charge 
excitations at each bond in Eq. (8). The result is\cite{alex,undr}
\beqn
A_\sigma (k,\omega)&=&Z\delta(\omega -\epsilon_{k\sigma})\nonumber \\
&+&
Z\sum_{l=1}^\infty \frac{(ln\frac{1}{Z})^l}{l!}\frac{1}{N}
\sum_{k'}[n_{k'}\delta(\omega+l\omega_0-\epsilon_{k'\sigma})
+(1-n_{k'})\delta(\omega-l\omega_0-\epsilon_{k'\sigma})]
\eeqn
which is easily seen to satisfy the sum rule
\beq
\int_{-\infty}^{\infty} d\omega A_\sigma (k,\omega)=1
\eeq
and describes the transfer of spectral weight from high to low frequencies as the
quasiparticle weight $Z$ (Eq. (38) increases when spin polarization develops. 
Similarly, the optical sum rule states
\beq
\int_0^{\omega_m}d\omega\sigma_1(\omega)=\frac{\pi e^2n}{2m^*}
\eeq
for the intra-band spectral weight of the optical conductivity
$\sigma_1(\omega)$. In Eq. (47), $\omega_m$ is a high  frequency cutoff that
excludes transitions to other bands. 
Using Eq. (43),\beq
\int_0^{\omega_m}d\omega\sigma_1(\omega)=\frac{\pi e^2n}{2m_0}Z(T,m)
\eeq
so that as the system becomes more coherent with increasing $Z$, spectral
weight is also transfered into the intra-band part of the optical
conductivity.
If Eq. (47) is integrated to infinity however
\beq
\int_0^{\infty}d\omega\sigma_1(\omega)=\frac{\pi e^2n}{2m^*}+
\int_{\omega_m}^\infty \sigma_1^{incoh}(\omega)=
\frac{\pi e^2n}{2m_0}
\eeq
with $m_0$ the bare mass. Thus, the extra spectral weight that
goes into intraband optical absorption has to be compensated
by a corresponding decrease in the incoherent contribution so
as to leave Eq. (49) invariant.

In the presence of a magnetic field the quasiparticle energies are
\beq
\epsilon_\sigma(\epsilon)=(1-2jI)\epsilon-\sigma(\frac{U+Jz}{2}m+Dh)-\mu
\eeq
with  $h$ a dimensionless magnetic field. Increasing $h$ gives rise to
increasing magnetization and decreasing bond charge $I$, as seen in Fig. 3. Hence, the
quasiparticle weight increases and the effective mass decreases. The 
magnetoresistance in this model is given by
\beq
\frac{\Delta\rho}{\rho}=\frac{\rho(h)-\rho(0)}{\rho(0)}=
2j\frac{I(T,m(h))-I(T,m(0))}{1-2jI(t,m(h))}
\eeq
and its behavior with temperature and magnetic field resembles that
seen in ferromagnets\cite{ferro}.
From the Drude form for the intra-band optical conductivity
\beq
\sigma_1(\omega)=\frac{ne^2}{m^*}\frac{\tau}{1+\omega^2\tau^2}=
\frac{ne^2}{m_0}Z(T,m)\frac{\tau}{1+\omega^2\tau^2}
\eeq
we conclude that the intra-band conductivity will increase with
application of a magnetic field, and correspondingly the high
frequency conductivity from incoherent processes will decrease.
To model both parts of the conductivity we take as a simple Ansatz
the spectral density Eq. (45) ignoring the momentum dependence
\beq
\sigma_1(\omega)=\frac{ne^2}{m_0}Z(T,m)[\frac{\tau}{1+\omega^2\tau^2}
+\frac{\pi}{2} \sum_{l=1}^\infty
\frac{(ln\frac{1}{Z})^l}{l!}\delta (\omega-l\omega_0)]
\eeq
which properly satisfies the sum rules Eqs. (48) and (49).
In figure 4 we show examples of the behavior expected under variation of
magnetic field and temperature. Qualitatively similar behavior is
seen in the optical properties of colossal magnetoresitive 
manganites\cite{manganites}
and of europium hexaboride\cite{hexaborides}.

Similarly, we expect the enhanced coherence in the ferromagnetic state
to be displayed in angle-resolved photoemission experiments:
under application of a magnetic field or lowering the temperature in
the ferromagnetic state, quasiparticle peaks should become stronger
reflecting the enhanced quasiparticle weight $Z(T,m)$. We will present
quantitative analysis and comparison with experiment elsewhere.

\section{Puzzles with the optical sum rule}
The optical sum rule in tight binding models needs to be treated
with some care. Consider the quasiparticle Hamiltonian Eq. (29) .
According to the discussion in the previous section, within mean field theory
Eq. (42) gives rise to a lowering of effective mass as spin polarization
develops, hence to an increased intra-band optical spectral weight according to
Eq. (47).

However, the polarization operator on the lattice is given by\cite{mahan2}
\beq
\vec{P}=e\sum_i\vec{R}_in_i
\eeq
with $R_i$ the position vector for site $i$. The current operator 
(in direction $\delta$) is 
obtained from its time derivative
\beq
J_\delta=\frac{dP_\delta}{dt}=\frac{i}{\hbar}[H,P_\delta]
\eeq
and is easily seen to be $independent$ $of$ $J$, because the exchange term
in Eq. (29) carries no current. Hence the $exact$ intra-band optical
sum rule for this Hamiltonian is\cite{maldague}
\bmath
\beq
\int_0^{\omega_m} d\omega \sigma_1(\omega)=
\frac{\pi a^2 e^2}{2 \hbar^2}<-T_t^\delta>
\eeq
\beq
<-T_t^\delta>=t\sum<\tilde{c}_{i\sigma}^\dagger \tilde{c}_{i+\delta\sigma}+h.c.>
\eeq
\emath
Eq. (56) predicts that as spin polarization develops and the
bond charge decreases the intra-band spectral weight will $decrease$.
This qualitatively contradicts the prediction of mean field theory for this very
same model, as well as the expectation based on the physical
considerations of the previous section. We are thus led to the remarkable
conclusion that the exact solution of the model Eq. (29) does worse than
its mean field solution in capturing essential aspects of its physics.

The situation can be remedied to some extent by including the
pair hopping term in the Hamiltonian. That term does carry a
current, and the sum rule Eq. (56) becomes\cite{color}
\bmath
\beq
\int_0^{\omega_m} d\omega \sigma_1(\omega)=
\frac{\pi a^2 e^2}{2\hbar^2}[<-T^t_\delta>+4<-T_\delta^{J'}>]
\eeq
\beq
<-T_\delta^{J'}>=-J'\sum_{i,\sigma}
<\tilde{c}_{i+\delta,\sigma}^\dagger \tilde{c}_{i+\delta,\sigma'}^\dagger
\tilde{c}_{i\sigma '}^\dagger \tilde{c}_{i\sigma }>
\eeq
\emath
In the normal state the expectation value Eq. (57b) is negative
and as spin polarization develops it will decrease in
magnitude leading to an increase in the optical spectral weight, in
accordance with qualitative expectation. These considerations illustrate
that the optical sum rule places severe constraints on what acceptable
effective low energy Hamiltonians are and on the relative magnitude
of their parameters. The subject clearly needs further investigation
which is outside the scope of this paper.

\section{Conclusions}

This paper started from the assumption that the electron creation
operator at site $i$ can be represented as
\bmath
\beq
c_{i\uparrow}^\dagger
=[1-(1-S)\tilde{n}(bond)]\tilde{c}_{i\uparrow}^\dagger
+incoherent\ part
\eeq
where the 'incoherent part' contains excitations of the local bond charge,
and $S$ describes the overlap of the bond-charge configuration ground state in the
presence and absence of the $\uparrow$ electron at site $i$. Similarly,
the work of I was based on the assumption that the operator can be
represented as
\beq
c_{i\uparrow}^\dagger=[1-(1-S)\tilde{n}(site)]\tilde{c}_{i\uparrow}^\dagger
+incoherent\ part
\eeq
More generally then we conclude that a general representation should be
\beq
c_{i\uparrow}^\dagger=[1-(1-S)\tilde{n}(local)]\tilde{c}_{i\uparrow}^\dagger
+incoherent\ part
\eeq
\emath
where $\tilde{n}(local)$ includes nearby site and bond charges. We have shown in I
that Eq. (58b) leads to superconductivity, and here that Eq. (58a) leads to
ferromagnetism. More generally, Eq. (58c) will lead to a unified description
of superconductivity and ferromagnetism where one or the other (or neither)
will dominate depending on characteristics of the system such as nature of
the orbitals, lattice structure and band filling.

In connection with superconductivity we note that in I the nature of the
boson degree of freedom was left somewhat unspecified. The analogy with
the situation discussed here makes it unambiguous that the boson degree of
freedom there is also electronic, as described e.g. by the electronic model with
two orbitals per site\cite{twoorb}, rather than e.g. a high frequency phonon.

The physics of ferromagnetism that results from Eq. (58a), and that of
superconductivity that results from Eq. (58b), are remarkably alike:
onset of the ordered state leads to increased quasiparticle coherence.
Spectral weight, both in the one-particle properties such as photoemission,
and in two-particle properties such as optical absorption, is transfered from
the incoherent high frequency background to the low frequency coherent
response. Quasiparticles undress, and become more like free particles, as
the ordered state develops.

Remarkably, the quasiparticle Hamiltonians that result from Eqs. (58a) and
(58b) lead respectively to the various off-diagonal matrix elements
of the Coulomb interaction in a local representation\cite{schrieffer2}: 
Eq. (58a) to exchange and
pair hopping (or equivalently bond-bond-charge repulsion), and Eq. (58b) to
correlated hopping (or bond-site charge repulsion). In our earlier work
we had shown that these off-diagonal matrix elements lead to metallic
ferromagnetism\cite{ferro} and hole superconductivity\cite{super}
 respectively, and that the common
aspect of the physics of both instabilities induced by these interactions
is effective mass reduction, which through the optical sum rule
implies transfer of optical
spectral weight to low frequencies. However, thinking about those interaction
terms as simply derived from static matrix elements of the Coulomb interaction
does not lead to an understanding of the optical spectral weight transfer process,
nor to the understanding that spectral weight transfer should also occur
in the one-particle Green's function.
Instead, the point of view presented in this paper and in I does.

What is the evidence that the physics discussed here takes place in
ferromagnetic metals? We have already mentioned that at least in some
ferromagnetic metals\cite{manganites,hexaborides}
 there is evidence for optical spectral weight
transfer from high to low frequencies as the ferromagnetic state
develops, either by decreasing the temperature or by increasing the
magnetic field\cite{henning}. 
We are not aware that evidence for undressing physics has been seen
yet in the single particle spectral function of ferromagnets, as would
be detected in photoemission experiments, but expect that it should  be
observable at least in manganites and hexaborides. For ferromagnets
that are more 'metallic' in the normal state, i.e. have higher
coherence, it will be more difficult to detect these effects since the
changes should be comparatively smaller. 

As suggested earlier\cite{ferro}
we believe the universal properties of negative magnetoresistance
of ferromagnets and anomalously large decrease of resistivity below
$T_c$ are evidence for effective mass reduction due to 'undressing', as
described by this paper, rather than for reduction of spin disorder
scattering as usually assumed\cite{friedel}. The difference is $not$ semantics:
in the Drude form for the optical conductivity, Eq. (52), changes
in $\tau$ and in $m^*$ will lead to different behavior at non-zero
frequencies, and it should be possible to decide this question
experimentally.

We also mention that another argument in favor of the picture of
ferromagnetism discussed here and in our earlier work is the
anomalous thermal expansion seen in ferromagnets below $T_c$.
This is clearly due to reduction of the $bond$ $charge$ as the systems
become ferromagnetic\cite{janak}, and points to the importance of the bond charge
in the phenomenon of ferromagnetism as described by our theory.

We believe that an experimental effort to detect the existence of
undressing in itinerant electron systems that become ferromagnetic
should be undertaken. There will of course be many system where
such evidence may be too small to be experimentally detectable.
However, if the evidence is found in a variety of different
itinerant ferromagnets it will provide convincing evidence that the
universal principle governing the transition to ferromagnetism in metals is
undressing. The fact that it may be possible to also understand
superconductivity with the same physical principle\cite{undr} lends further
support to the possibility that 'undressing' may capture the essential
physics of both phenomena.

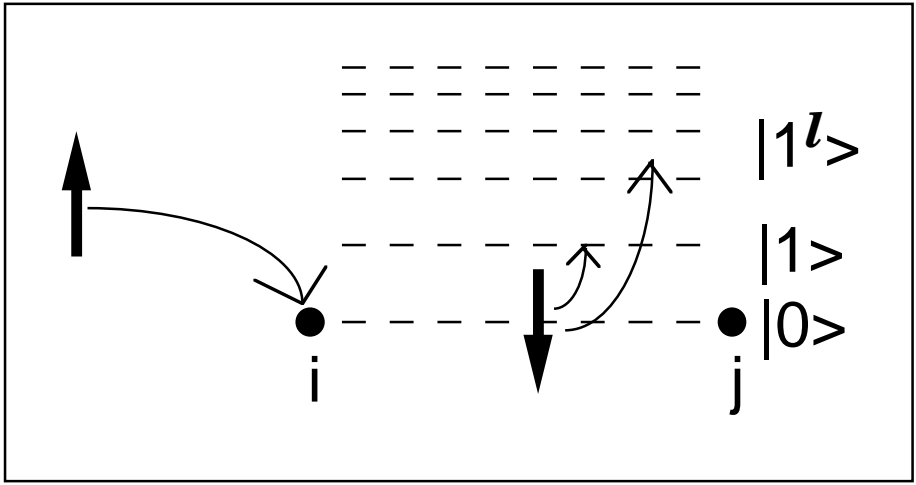
\begin{figure}
\caption {
When the spin $\uparrow$ electron is created at site $i$, the state of the
bond charge between sites $i$ and $j$, represented here by a $\downarrow$-spin electron,
changes because of Coulomb repulsion. It may make a diagonal transition to its 
new ground state $|1>$ (coherent process) or be left in an excited state 
$|1^l>$ (incoherent process).}
\label{Fig. 1}
\end{figure}
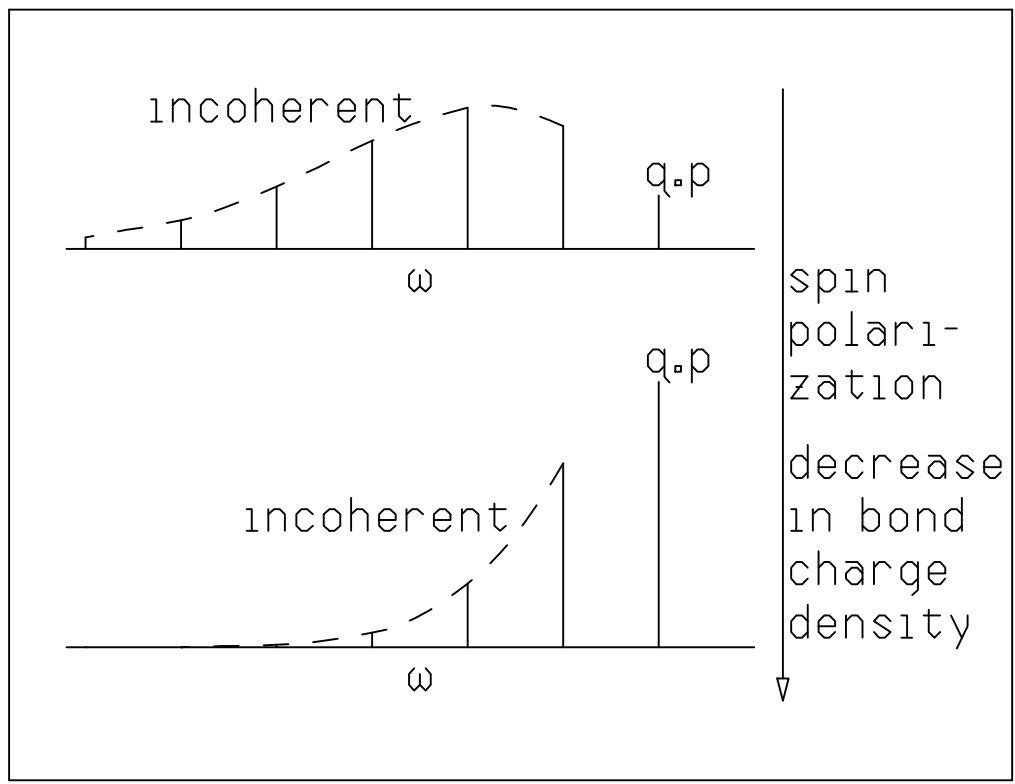
\begin{figure}
\caption {Schematic picture of the single particle spectral function resulting
from the physics proposed here. In the unpolarized state (upper part) there is a large 
electronic bond charge density, most of the spectrum is incoherent and
the quasiparticle weight $Z$, the height of the peak labeled q.p., is small.
 As spin polarization develops (lower part)
the bond charge density is reduced, the quasiparticle weight increases and
spectral weight in the incoherent part of the spectrum becomes smaller. 
For the example in the figure, $Z=0.1$ for the upper part and $Z=0.5$ for
the lower part. Such qualitative features (albeit less extreme) should
be seen in photoemission spectra of ferromagnetic metals.
}
\label{Fig. 2}
\end{figure}
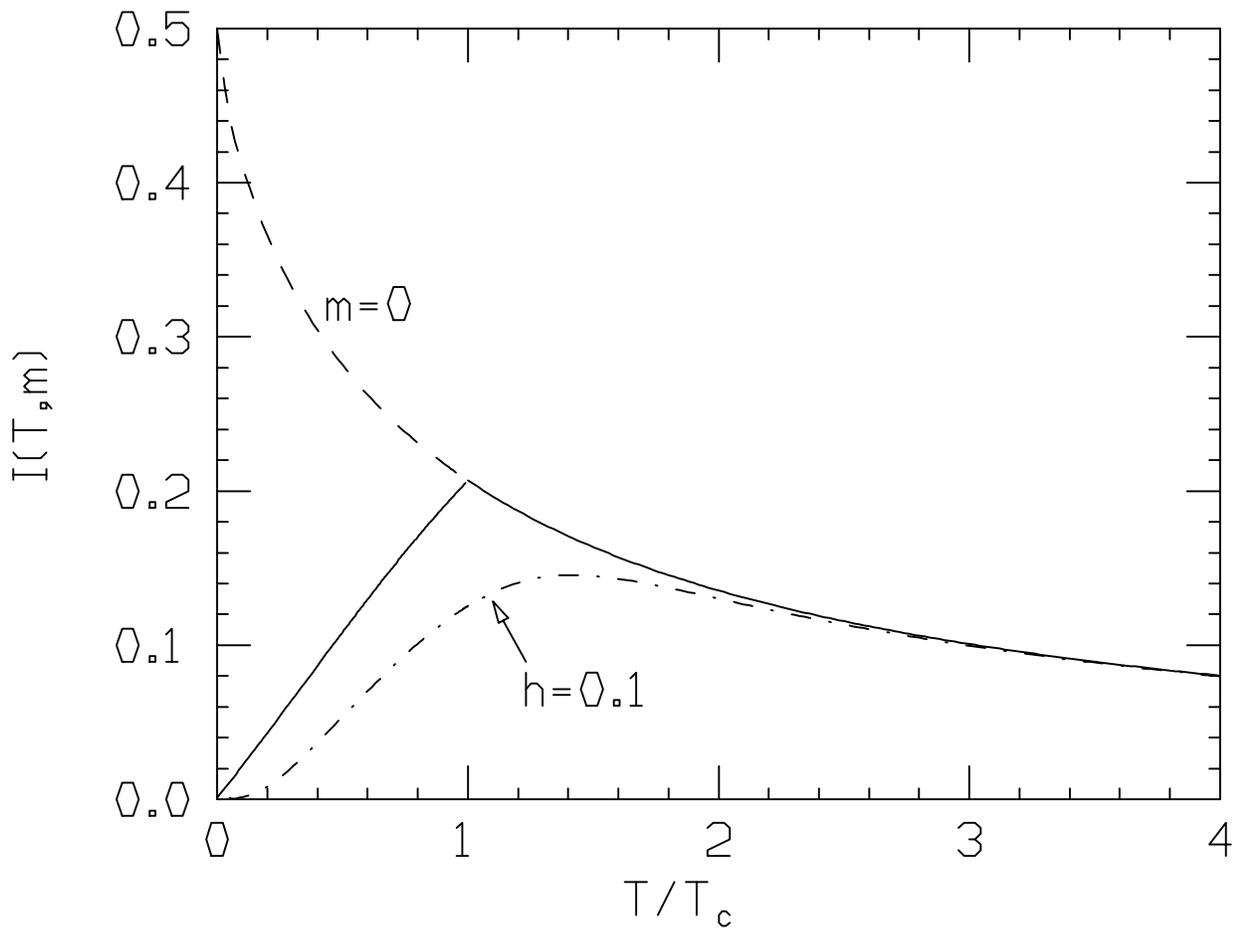
\begin{figure}
\caption { Characteristic behavior of the bond charge $I(T,m)$. Parameters
used here are $j=1$, $u=0$,  for $n=1$. As the temperature is lowered above
$T_c$ the bond charge increases, and when spin polarization develops below
$T_c$ it decreases again. If no spin polarization is allowed to 
develop (dashed curve labeled $m=0$) the bond charge continues increasing
as $T$ is lowered, and correspondingly the effective bandwidth decreases,
at a cost in kinetic energy. That cost is relieved by the development
of spin polarization. In the presence of a magnetic field (Eq. (50))
(dash-dotted line) the magnetization increases and the bond charge is
reduced.
}
\label{Fig. 3}
\end{figure}
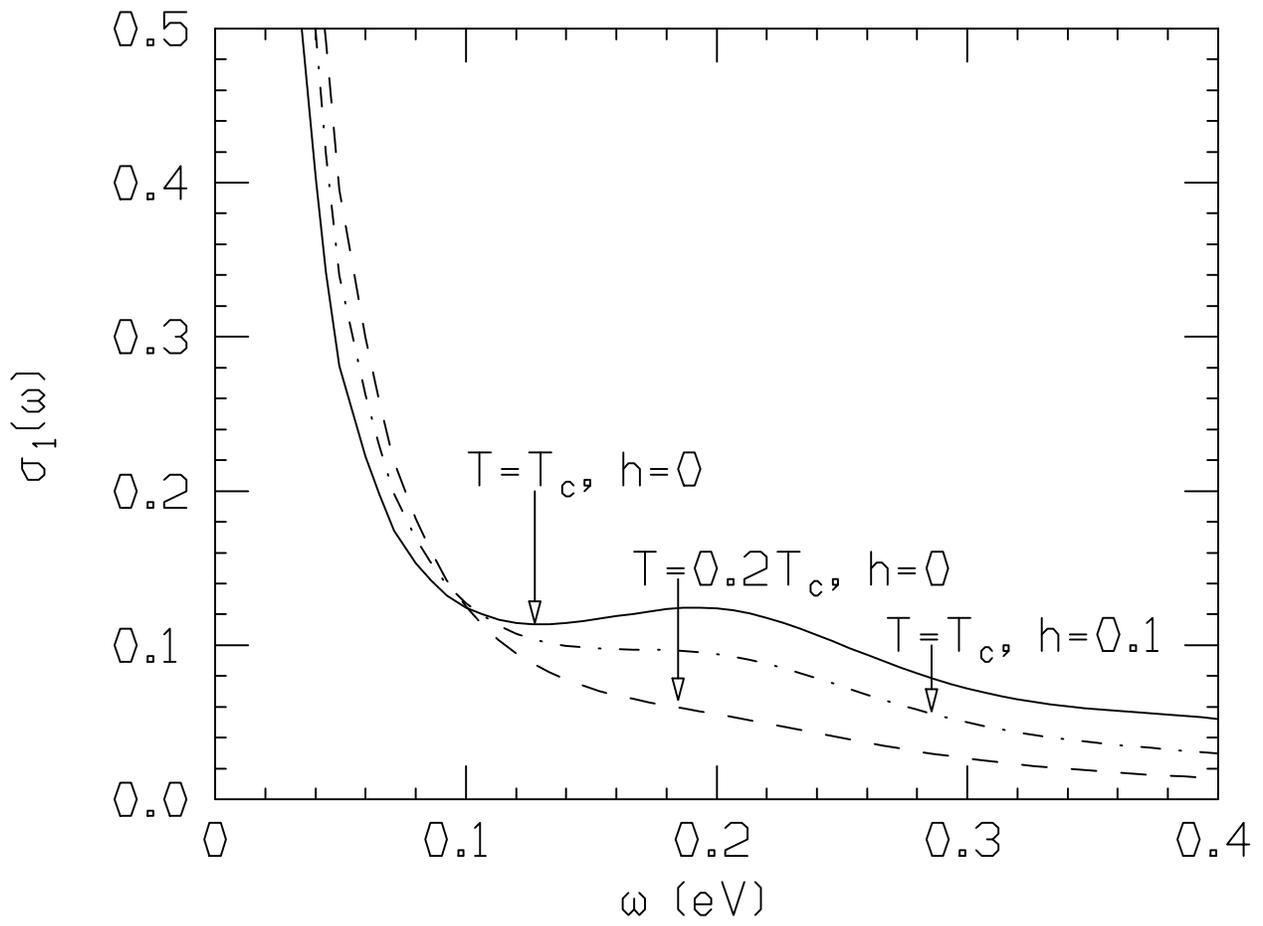
\begin{figure}
\caption {Characteristic behavior expected for the real part of
the optical conductivity $\sigma_1(\omega)$ (arbitrary units), for
the same parameters as in Fig. 3, with $D=1eV$. The expression Eq. (53) is used,
with $\omega_0=.2eV$ and $\tau=50 eV^{-1}$. The $\delta$-functions
in Eq. (53) were broadened to Lorentzians with half-width $\Gamma=0.1eV$.
Full line shows optical absorption at $T_c$ in the absence of a magnetic
field. Both when the temperature is lowered (dashed line) and when a
magnetic field is applied (dash-dotted line) spectral weight is transfered
from the incoherent region to the low frequency Drude region due to 
the undressing induced by spin polarization.
}
\label{Fig. 4}
\end{figure}

\end{document}